\documentclass[conference]{IEEEtran}
\IEEEoverridecommandlockouts
\usepackage{amsmath}
\usepackage{amsthm}
\usepackage{mathrsfs}
\usepackage{graphicx}
\usepackage{cite}

\ifCLASSINFOpdf
\else

\fi

\begin{document}
\title{Joint Network and Gelfand-Pinsker Coding for 3-Receiver Gaussian Broadcast Channels with Receiver Message Side Information}
\author{\IEEEauthorblockN{Jin Sima and Wei Chen}
\IEEEauthorblockA{Tsinghua National Laboratory for Information Science and Technology (TNList)\\}
\IEEEauthorblockA{Department of Electronic Engineering, Tsinghua University, Beijing, CHINA\\
Email: smj13@mails.tinghua.edu.cn, wchen@tsinghua.edu.cn
}\thanks{This paper is partially supported by the National Basic Research Program of China (973 Program 2013CB336600), NSFC Excellent Young Investigator Award No. 61322111, MoE new century talent program under No. NCET-12-0302, Beijing Nova Program No. Z121101002512051, and National Science and Technology Key Project No. 2013ZX03003006-005.}%
}
\maketitle
\begin{abstract}
The problem of characterizing the capacity region for Gaussian broadcast channels with receiver message side information appears difficult and remains open for $N\ge 3$ receivers. This paper proposes a joint network and Gelfand-Pinsker coding method for 3-receiver cases. Using the method, we establish a unified inner bound on the capacity region of 3-receiver Gaussian broadcast channels under general message side information configuration. The achievability proof of the inner bound uses an idea of joint interference cancelation, where interference is canceled by using both dirty-paper coding at the encoder and successive decoding at some of the decoders. We show that the inner bound is larger than that achieved by state of the art coding schemes. An outer bound is also established and shown to be tight in 46 out of all 64 possible cases.
\end{abstract}
\IEEEpeerreviewmaketitle

\section{Introduction}
A broadcast channel with receiver message side information represents a scenario where a sender wishes to communicate messages to the receivers and each receiver knows part of the messages a priori. Such scenario arises in several contexts, e.g., downlink transmission of decode and forward schemes in a multiway relay channel \cite{gunduz2009multi}.

Many explorations have been made in the past decade in order to find the capacity region for broadcast channels with receiver side information. The capacity region for 2-receiver Gaussian broadcast channels with receiver side information is well known. It is found in \cite{wu2007broadcasting},\cite{kramer2007capacity} that when the weaker receiver knows a priori the message intended for the stronger receiver, the capacity region can be enlarged using network coding or lattice coding. When the weaker receiver observes no message side information, the presence of message side information will not help increase the capacity region.

Characterizing the capacity region for multi-receiver Gaussian broadcast channels under general side information configuration appears difficult and remains open. 
Special cases are investigated in \cite{xie2007network} where each receiver knows a priori the messages intended for all other receivers. For more general cases, coding schemes are proposed. The state of the art schemes include network coding with time sharing \cite{gunduz2009multi} and separate physical and network coding \cite{liu2009gaussian,chen2006cross}, which applies index coding \cite{bar2011index} in the network layer and superposition coding in the physical layer. The coding schemes in \cite{liu2009gaussian,chen2006cross} can be viewed as successive network coding and interference cancelation. However, they are generally suboptimal since separation of network and channel coding leads to limited performance of network coding and suffers from capacity loss. 

In this paper we consider 3-receiver Gaussian broadcast channels with receiver message side information. We propose a joint network and Gelfand-Pinsker coding method, which embeds network coding into Gelfand-Pinsker coding \cite{Pinsker1983writing}. The coding method provides a unified coding structure for all message side information configurations. With this method, we introduce a joint interference cancelation technique and then derive a unified inner bound on the capacity region of 3-receiver Gaussian broadcast channels under general side information configuration. The inner bound is larger than that achieved by time sharing and separate network and physical coding. An outer bound is also established and shown to be tight in 46 out of all 64 possible cases. We note that in \cite{asadi2014awgn}, similar results on the capacity regions of 3-receiver Gaussian broadcast channels with receiver side information are presented. However, coding techniques of this paper and \cite{asadi2014awgn} are different. Moreover, all cases of side information configuration are considered in our work.



\section{System Model}\label{system model}
Consider a 3-receiver Gaussian broadcast channel as depicted in Fig.1, where the outputs are given by
\begin{equation}\label{gaussianbc}
Y_i = X + Z_i, ~~~~i=1,2,3.
\end{equation}
The input $X$ satisfies an average power constraint of $P$ and
the noise components $Z_i\sim \mathcal{N}(0,N_i)$ are independent. Without loss of generality, it is assumed that $N_1<N_2<N_3$.

The sender wishes to send a set of independent messages $\mathcal{W}=\{W_1,W_2,W_3\}$ to the receivers, where $W_i$ is intended for receiver $i$. Receiver $i$ observes $Y_i$ and knows a priori a subset of messages $\overline{\mathcal{W}}_i\subset\mathcal{W}$. The configuration of message side information can be characterized by a routing matrix $\boldsymbol{A}$ \cite{chen2006cross}, where $\boldsymbol{A}$ is
a $3\times 3$ matrix with elements
\renewcommand\arraystretch{1}
$$ a_{ij}=\left\{
\begin{array}{rcl}\label{routing}
1       && {\text{if Receiver i knows }W_j\text{ a priori},}\\
0       && {\text{else}.}
\end{array} \right. $$
Note that receiver $i$ does not know $W_i$ a priori, hence $a_{ii}=0$.

The definitions of a $(2^{nR_1},2^{nR_2},2^{nR_3},n)$ code, achievable rates and capacity region follow a standard way as in \cite{el2011network}, which are omitted in this paper due to the page limitation.
%
%
For convenience, throughout this paper, we denote $\mathcal{I}=\{1,2,3\}$ as the set of all receivers.

\begin{figure}\label{fig:1}
\center
\includegraphics[scale=0.66]{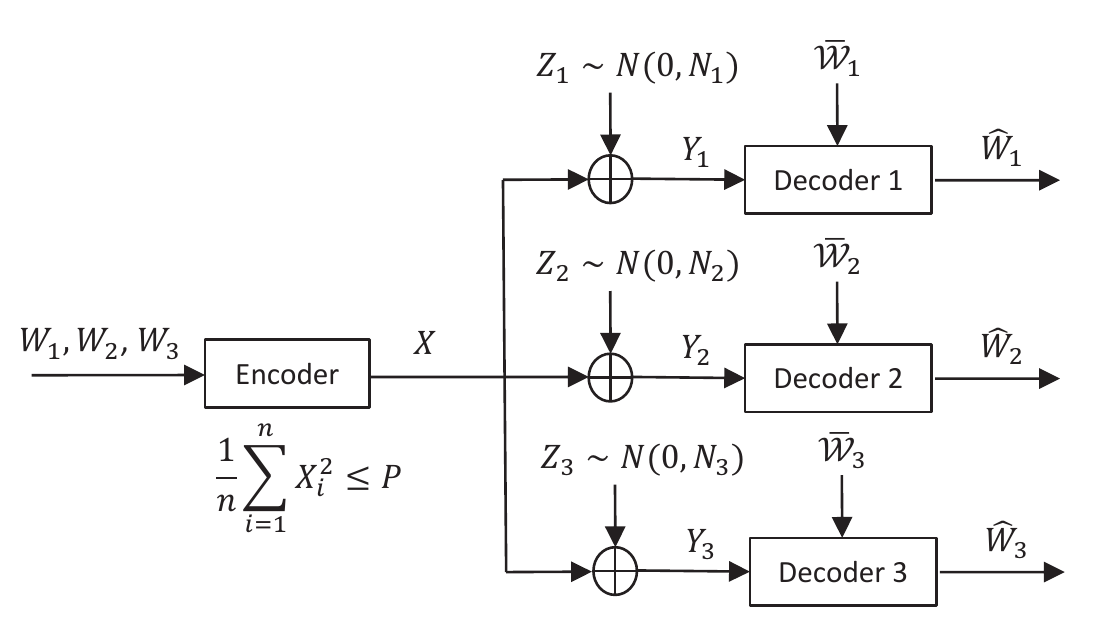} 
\caption{3-receiver Gaussian broadcast channels with receiver side information}
\setlength\belowcaptionskip{0pt}
\end{figure}
\section{Joint Network and Gelfand-Pinsker Coding}
Consider a 3-receiver memoryless BC with state
\begin{equation}\label{statebc}
(\mathcal{X}\times \mathcal{S},p(y_1,y_2,y_3|x,s)p(s),\mathcal{Y}_1 \times \mathcal{Y}_2 \times \mathcal{Y}_3),
\end{equation}
with input $X \in \mathcal{X}$, outputs $Y_i \in \mathcal{Y}_i$, and state $S \in \mathcal{S}$. The sender wishes to send messages $\{W_1,W_2,W_3\}$ to all receivers, where each receiver knows messages $\overline{\mathcal{W}}_i$ a priori. The routing matrix $\boldsymbol{A}$ is similarly defined as in the previous section. The state sequence $S^n$ is available noncausally at the encoder and some of the decoders, where the availability at the decoders is indicated by a tuple $(b_1,b_2,b_3)$. Set $b_i=1$ if the sequence $S^n$ is available at decoder $i$ and $b_i=0$ otherwise.

\textbf{\emph{Definition 1:}} A set of receivers $\mathcal{V} \subset \mathcal{I}$ is \emph{acyclic} if there does not exist a subset $ \{i_m:m=1,\ldots,M\}\subset \mathcal{V}$ such that $a_{i_mi_{m+1}}=1$ for $m=1,\ldots,M$ (Here $i_{M+1}=i_1$).

Here we note that in particular, an empty set or a set with one element is acyclic. Denote $\mathcal{L}_{\boldsymbol{I}}=\{\mathcal{V}|\mathcal{V}\subset \mathcal{I},~\mathcal{V}$ is acyclic $\}$. The joint network and Gelfand-Pinsker coding scheme is presented as follows.

\textbf{\emph{Theorem 1:}} For the broadcast channel \eqref{statebc}, for any $\mathcal{J}\subset\mathcal{I}$, a rate tuple  $(R_1,R_2,R_3)$ is achievable if
\begin{equation}\label{codingblockcapacity}
\begin{split}
&\max_{\mathcal{V}\in\mathcal{L}_{\boldsymbol{I}}}\sum_{j\in \mathcal{V},j\notin\mathcal{O}_i}R_j \le C_i,~~~~i\in \mathcal{J};\\
&R_i=0,~~~~i\notin\mathcal{J}
\end{split}
\end{equation}
for some $p(u|s),f(u,s)$, where $\mathcal{O}_i=\{j:a_{ij}=1\}$ and $C_i=\max\{b_iI(U;Y_i|S),I(U;Y_i)-I(U;S)\}$.
\begin{IEEEproof}
We prove the result for the following cases respectively. 1) There exists a pair $(i,j)$ which satisfies $a_{ij}=a_{ji}=1$. 2) There does not exist such a pair.
\\\emph{Case~$1$:} There exists a pair $(i,j)$, where $a_{ij}a_{ji}=1$.
 Without loss of generality, we assume that $a_{12}a_{21}=1$. Let $R=$ $\max\{R_{1},R_{2}\}$.
For a message triple $(w_1,w_2,w_3)$, where $w_i\in \{1,\ldots,2^{nR_i}\}$, define the index function
\begin{equation}
k(w_1,w_2,w_3) \buildrel \Delta \over =  (w_3-1)2^{nR}+w_1 \oplus w_2.
\end{equation}
Here $w_1 \oplus w_2 \buildrel \Delta \over = (w_1+w_2)~mod~(2^{nR})$.
\\\emph{Codebook~Generation:} Fix $p(u|s)$ and $x(u,s)$ that satisfy \eqref{codingblockcapacity}. Independently generate $2^{n(R_3+R)}$ subcodebooks $\mathcal{C}(m)$, $m \in \{1,\ldots,2^{n(R_3+R)}\}$, each consists of $2^{n\widetilde{R}}$ sequences $u^n(l),~l \in \{(m-1)2^{n\widetilde{R}}+1,\ldots,m2^{n\widetilde{R}}\}$, randomly and independently generated according to $\prod_{i=1}^{n}p(u_i)$.
\\\emph{Encoding:} Given the state $s^n$ and the message tuple $(w_1,w_2,w_3)$, where $w_i\in\{1,\ldots,2^{nR_i}\}$, the encoder chooses a sequence $u^n(l) \in \mathcal{C}(k(w_1,w_2,w_3))$ such that $(u^n(l),s^n) \in A_{\epsilon'}^{(n)}$. If there is no such one, it chooses $u^n(1)$. Then the encoder transmits $f_i=f(u_i,s_i)$ at time $i=1,\ldots,n$.
\\\emph{Decoding:} Let $\epsilon>\epsilon'$, Based on the received $Y_i^n$, decoder $i$ looks for the unique sequence $u^n(l)$ that satisfies $(u^n(l),Y_i^n) \in A_{\epsilon}^{(n)}$ in subcodebooks $\mathcal{C}(k(w_1,w_2,w_3))$, where the tuple $(w_1,w_2,w_3)$ has the same $\overline{w}_i$. If there is exactly one such sequence in some subcodebook $\mathcal{C}(k(w'_1,w'_2,w'_3))$, decoder 3 declares that $w'_3=\lfloor \frac{k(w'_1,w'_2,w'_3)}{2^{nR}}\rfloor+1$ is sent, decoders $1$ and $2$ declare that $(k(w'_1,w'_2,w'_3)-w'_2)~mod~(2^{nR})$ and $(k(w'_1,w'_2,w'_3)-w'_1)~mod~(2^{nR})$ are sent respectively. Otherwise, decoder $i$ declares an error.
\\\emph{Analysis of the probability of error:}
Without loss of generality, assume that the message tuple is $(1,1,1)$ and the chosen sequence is $u^n(1) \in \mathcal{C}(k(1,1,1))$. For decoder $i\notin\mathcal{J}$, the probability of error can be ignored since $R_i=0$. For decoder $i\in \mathcal{J}$, an error is attributed to the following events

$E_{1i} = \{(u^n(l),s^n) \notin A_{\epsilon'}^{(n)}~\text{for all }u^n(l) \in \mathcal{C}(k(1,1,1))\}$

$E_{2i} = \{(u^n(1),Y_i^n) \notin A_{\epsilon}^{(n)}\}$

$E_{3i} = \{(u^n(l),Y_i^n) \in A_{\epsilon}^{(n)}~\text{for some }l\ne1~and~u^n(l) \in$

$\mathcal{C}(k(w_1,w_2,w_3)),~\text{where the tuple}~(w_1,w_2,w_3)$ has the

same $\overline{w}_i\}.$

The probability of error can be upper bounded as
\begin{equation}\label{errorbound}
P^{(n)}_{ei} \le P(E_{1i})+P(E_{1i}^c \cap E_{2i})+P(E_{3i}).
\end{equation}
Since sequences $u^n(l),~l=1,\ldots,2^{n(R+R_3+\tilde{R})}$, are generated independently of $s^n$,
$P(E_{1i})$ tends to $0$ as $n\rightarrow \infty$, if $\widetilde{R} > I(U;S)+\delta'$ (see \cite{el2011network} for details). Moreover, by the asymptotic equipartition
property, $P(E_{1i}^c \cap E_{2i})\rightarrow 0$ as $n \rightarrow \infty$.

Finally, denote $R_i^{sum}=\max_{\mathcal{V}\in\mathcal{L}_{\boldsymbol{I}}}\sum_{j\in \mathcal{V},j\notin\mathcal{O}_i}R_j$.
Then with the same $\overline{w}_i$ fixed, there are $2^{nR_i^{sum}}$ different subcodebook indices $k(w_1,w_2,w_3)$. Observe that sequences $u^n(l)$ are generated independently of each other and of $Y_i^n$, we have
\begin{equation}\label{e3i}
\begin{split}
P(E_{3i})&\le 2^{nR_i^{sum}}2^{n\widetilde{R}}P((u^n(l),Y_i^n) \in A_{\epsilon}^{(n)})\\
&\le 2^{n(R_i^{sum} + \widetilde{R} - n(I(U;Y_i)-\delta)},
\end{split}
\end{equation}
where the second inequality follows by the fact that $P((u^n(l),Y_i^n) \in A_{\epsilon}^{(n)})\le 2^{-n(I(U;Y_i)-\delta)}$ \cite{el2011network}. Therefore, $P(E_{3i}) \rightarrow 0$ as $n\rightarrow \infty$, if $R_i^{sum} \le I(U;Y_i)-\delta - \widetilde{R}$.

Combining these results, we conclude that $P^{(n)}_{ei}$ tends to zero as $n \rightarrow \infty$, if $\max_{\mathcal{V}\in\mathcal{L}_{\boldsymbol{I}}}\sum_{j\in \mathcal{V},j\notin\mathcal{O}_i}R_j <I(U;Y_i)-I(U;S) - \delta-\delta'$. When $b_i=1$, i.e., the state sequence $s^n$ is known at decoder $i$, it can be incorporated into $Y_i^n$ \cite{steinberg2005coding}. By substituting $Y_i \buildrel \Delta \over = (Y_i,S)$, the term $I(U;Y_i)-I(U;S)$ becomes $I(U;Y_i|S)$. Hence $P^{(n)}_{ei}$ tends to zero as $n \rightarrow \infty$, if $\max_{\mathcal{V}\in\mathcal{L}_{\boldsymbol{I}}}\sum_{j\in \mathcal{V},j\notin\mathcal{O}_i}R_j <I(U;Y_i|S) - \delta-\delta'$.
\\\emph{Case 2:} There does not exist a pair $(i,j)$ where $a_{ij}=a_{ji}=1$. In this case, similar argument can be applied
 if we define the index function
\begin{equation}
k(w_1,w_2,w_3) \buildrel \Delta \over =  (w_3-1)2^{n(R_1+R_2)}+(w_2-1)2^{nR_1}+w_1
\end{equation}
and generate $2^{n(R_1+R_2+R_3)}$ subcodebooks $\mathcal{C}(m)$ similarly in the codebook generation step. Thus the theorem is proved.
\end{IEEEproof}
\textbf{\emph{Remark 1:}} Note that decoder $i\in \mathcal{J}$ is able to decode the full sequence $u^n(l)$.

\section{Inner and Outer Bounds on Capacity Region}
In this section we use joint network and Gelfand-Pinsker coding to establish an inner bound to the capacity region for channel \eqref{gaussianbc}. Then we provide an outer bound and show that it is tight in most of the cases.

%

\subsection{Inner Bound Achieved by Joint Network and Gelfand-Pinsker Coding}
We begin with some definitions that are needed to describe our coding scheme.

\textbf{\emph{Definition 2:}}
A set $\mathcal{V}\subset\mathcal{I}$ is \emph{complete} if for every pair $i,j\in \mathcal{V}$ with $i<j$, we have $a_{ji}=1$. In particular, a set with one element is complete. Further, a complete set $\mathcal{V}$ is \emph{maximum} if for each element $j\in\mathcal{I}\backslash\mathcal{V}$, $\mathcal{V}\cup\{j\}$ is not complete.

Denote
$\mathcal{K}_{\boldsymbol{I}}=\{\mathcal{V}:\mathcal{V}\subset\mathcal{I}$ \text{ is maximum complete}$\}$. 
The following theorem provides an inner bound that can be achieved by joint network and Gelfand-Pinsker coding.

\textbf{\emph{Theorem 2:}}
Denote $\mathcal{R}_{in}$ as the set that consists of all rate tuples $(R_1,R_2,R_3)$ satisfying
\begin{equation}\label{capacityregion}
\sum_{k\in \mathcal{V}}R_k\le\sum_{l=1}^3C\Big(\frac{P_l}{\min_{i\in\mathcal{K}_l,i\in\mathcal{V}}N_i+\sum_{m<l}P_m}\Big)
\end{equation}
for all sets $\mathcal{V}$ such that $\mathcal{V}\cap\mathcal{K}_l$ is acyclic or empty, and for some nonnegative $(P_1,P_2,P_3)$ such that $\sum_{l=1}^3P_l=P$. Here
\begin{equation}
\mathcal{K}_l=\text{arg}\min_{\mathcal{K}\in\mathcal{\mathcal{K}_{\boldsymbol{I}}},\mathcal{K}\ni l}(\min\mathcal{K}+\max\mathcal{K}),~l=1,2,3.
\end{equation}
Then $\mathcal{R}_{in}$ forms an inner bound to the capacity region for Gaussian broadcast channel \eqref{gaussianbc}.
\begin{IEEEproof}
Let $d=a_{31}+a_{32}+a_{21}$. We consider the cases when $d=3,2,1,0$ respectively.
Case 1: $d=3$. In this case, we have $\mathcal{K}_{\boldsymbol{I}}=\{\{1,2,3\}\}$ and $\mathcal{K}_1=\mathcal{K}_2=\mathcal{K}_3=\{1,2,3\}$. The constraint \eqref{capacityregion} reduces to
\begin{equation}
\sum_{k\in \mathcal{V}}R_k\le C\Big(\frac{P}{\min_{i\in\mathcal{V}}N_i}\Big).
\end{equation}
The achievability proof follows from Theorem 1, by setting $\mathcal{J}=\{1,2,3\}$, $S=0$, and $U=X\sim\mathcal{N}(0,P)$ in \eqref{codingblockcapacity}.

Case 2: $d=2$. In this case, $\mathcal{K}_{\boldsymbol{I}}=\{\{k_1,k_2\},\{k_2,k_3\}\}$, where $k_1,k_2,k_3\in\mathcal{I}$ are different receivers. The encoder splits the messages $W_i,~i=1,2,3,$ into $3$ parts $W_{il},~l=1,2,3,$ at rate $R_{il}$, where $W_{il}=\emptyset$ and $R_{il}=0$ for $i\notin\mathcal{K}_l$. Then it transmits $X=x_1+x_2+x_3$, where $x_3,x_2$, and $x_1$ are generated sequentially as follows.

To generate $x_l$, the encoder considers $s_l=\sum_{j>l}x_j$ as noncausally known interference, maps the message tuple $(w_{1l},w_{2l},w_{3l})$ into a codeword $u_l$ and then computes $x_l=f_l(u_l,s_l)$. The encoder does these using joint network and Gelfand-Pinsker coding, where it sets $U_l$ as the dirty-paper coding auxiliary random variable $\alpha_lS_l+X_l$, with $X_l\sim\mathcal{N}(0,P_l)$. Here $\alpha_l=\frac{P_l}{N_{k_1}+\sum_{m\le l}P_m}$ if $k_1\in\mathcal{K}_l$ and $\alpha_l=\frac{P_l}{N_{k_3}+\sum_{m\le l}P_m}$ otherwise.

Decoders $k_1,k_3$ decode their messages $W_{k_jl},~j=1,3,~l=1,2,3$, by treating $\sum_{j<l}x_j$ as additional noise, while decoder $k_2$ does the following steps for $l=3,2,1$, sequentially.

\emph{Step 1}: Based on the observed interference $s_l$ $(s_3=0)$, decode $u_l$ by treating $\sum_{j<l}x_j$ as additional noise and then decode the message $W_{k_2l}$.

\emph{Step 2:} Compute $x_l=u_l-\alpha_{l}s_l$ and renew the observed interference $s_{l-1}=x_l+s_l$.

Substituting $\mathcal{J}=\mathcal{K}_l,$ $X=X_l$, $U=U_l,$ and $b_{k_2}=1$ in \eqref{codingblockcapacity} for $l=3,2,1,$ respectively, we conclude from Theorem 1 and Remark 1 that decoders $i$ decode the corresponding sequences $u_l$ and messages $W_{il}$ correctly if
\begin{equation}\label{constrainttuple}
\begin{split}
&\sum_{i\in\mathcal{V}_l}R_{il}\le C\Big(\frac{P_l}{\min_{j\in\mathcal{V}_l}N_j+\sum_{m<l}P_m}\Big),\\
&R_{il}=0,~~~~i\notin \mathcal{K}_l,\\
\end{split}
\end{equation}
for all acyclic sets $\mathcal{V}_l\in\mathcal{K}_l,~l=1,2,3$. 
Further substituting $R_i=\sum_{l=1}^3R_{il}$ in \eqref{constrainttuple} and using Fourier-Motzkin elimination to eliminate $R_{ij}$, we obtain inequalities of the form
\begin{equation}\label{FourierMotzkin}
\sum_{i\in\bigcup_{l\in\mathcal{T}}\mathcal{V}_l}R_{i}\le \sum_{l\in\mathcal{T}}C\Big(\frac{P_l}{\min_{j\in\mathcal{V}_l}N_j+\sum_{m<l}P_m}\Big)
\end{equation}
where $\mathcal{T}\subset\{1,2,3\}$, and $\mathcal{V}_l\bigcap\big(\bigcup_{i\in\mathcal{T},i\ne l}\mathcal{V}_i\big)\ne\emptyset$ if $|\mathcal{T}|\ge 2$. Let $\mathcal{V}=\bigcup_{l\in\mathcal{T}}\mathcal{V}_l$. We have $\mathcal{V}\cap\mathcal{K}_l\subset\mathcal{V}_l$. Then constraint \eqref{capacityregion} becomes \eqref{FourierMotzkin}. Hence the region $\mathcal{R}_{in}$ is achievable.


Case 3: $d=1$. If $a_{21}=1$ or $a_{31}=1$, then $\mathcal{K}_{\boldsymbol{I}}=\{\{1,j_1\},\{j_2\}\}$ ($j_1,j_2\in\{2,3\}$ are different). And we have $\mathcal{K}_1=\mathcal{K}_{j_1}=$ $\{1,j_1\}$ and $\mathcal{K}_{j_2}=j_2$. The encoder generates $X$ similarly as in case 2 except that $\alpha_1=\frac{P_1}{N_{j_1}+P_1}$, $\alpha_{j_1}=\frac{P_{j_1}}{N_{j_1}+\sum_{m\le {j_1}}P_m}$ and $\alpha_{j_2}=\frac{P_{j_2}}{N_{j_2}+\sum_{m\le j_2}P_m}$. Decoder $1$ follows similar steps as $k_2$ does in case 2 while decoders $j_1,j_2$ decode messages $W_{j_nl},~n=1,2,~l=1,2,3,$ by treating $\sum_{m<l}x_m$ as additional noise.
Note that $x_{j_2}$ is the dirty-paper coding signal for receiver $j_2$, decoder $1$ can also decode $u_{j_2}$ since it has stronger receiver. The rest of the proof follows similarly as in case 2.

If $a_{32}=1$, we have $\mathcal{K}_1=\{1\}$ and $\mathcal{K}_2=\mathcal{K}_3=\{2,3\}$. In this case, the input $X$ is similarly generated except that $\alpha_1=\frac{P_1}{N_1+P_1}$, $\alpha_2=\frac{P_2}{N_2+P_1+P_2}$ and $\alpha_3=\frac{P_3}{N_2+P_1+P_2+P_3}$. Decoders $2,3$ do steps $1$ and $2$ as $k_2$ does in case 2 for $l=3,2$, while decoder $1$ directly decodes message $W_{11}$. The achievability proof follows from Theorem 1.

Case 4: $d=0$. In this case, we have $\mathcal{K}_l=\{l\}$. The region \eqref{capacityregion} can be achieved by successive dirty paper coding \cite{caire2003achievable},
which is a special case of successive joint network and Gelfand-Pinsker coding. This completes the proof.
\end{IEEEproof}
\textbf{\emph{Remark 2:}} The achievability proof in case 2 - 3 employs both dirty-paper coding at the encoder and successive decoding at the decoders to cancel the inference. We refer to this technique as joint interference cancelation.

In general, the rate region achieved by combined network coding and interference cancelation schemes, with network coding performed in complete sets, is given by
\begin{equation}
\begin{split}
&\mathcal{C}(M,\boldsymbol{\mathcal{K}}(M))=\Big\{(R_1,R_2,R_3):~~\sum_{l=1}^MP_l=P,\\
&\sum_{k\in \mathcal{V}}R_k\le\sum_{l=1}^MC\Big(\frac{P_l}{\min_{i\in\mathcal{K}_l,i\in\mathcal{V}}N_i+\sum_{m<l}P_m}\Big),\\
&\forall \mathcal{V}\cap\mathcal{K}_l\subset \mathcal{K}_{\boldsymbol{I}}\cup\emptyset\Big\}.
\end{split}
\end{equation}
Here $\boldsymbol{\mathcal{K}}(M)=(\mathcal{K}_i:\mathcal{K}_i\subset\mathcal{I}$ is complete $,i=1,\ldots,M)$.
It can be shown that $\mathcal{R}_{in}$ characterizes the largest region of this kind. Specifically, we have the following proposition.

\textbf{\emph{Proposition 1:}}
\begin{equation}\label{maxweightedsumrate}
\mathcal{R}_{in}=Conv\{\bigcup_{M,\boldsymbol{\mathcal{K}}(M)}\mathcal{C}(M,\boldsymbol{\mathcal{K}}(M))\},
\end{equation}
where $Conv(C)$ is the convex hull of set $C$.
\begin{IEEEproof}
We briefly outline the proof as follows. Denote $\mathcal{C}$ as the RHS region of \eqref{maxweightedsumrate}. Then we have $\mathcal{R}_{in}\subset\mathcal{C}$. The proposition can be proved by showing that 1) $Conv(\mathcal{C})\subset Conv(\mathcal{R}_{in})$ and 2) $\mathcal{R}_{in}$ is convex. To prove 1) and 2), it suffices to show that the problem
\begin{equation}\label{optimization}
\max_{(R_1,R_2,R_3)\in\mathcal{C}}\mu_1R_1+\mu_2R_2+\mu_3R_3
\end{equation}
has a unique solution $(R_1,R_2,R_3)\in\mathcal{R}_{in}$ given any positive tuple $(u_1,u_2,u_3)$. To solve problem \eqref{optimization}, we adopt a similar utility function approach as in \cite{tse1998multiaccess}.

Note that when $a_{31}a_{32}a_{21}=1$, $\mathcal{R}_{in}$ and $\mathcal{C}$ reduce to the polytope $\mathcal{C}(1,\{1,2,3\})$. We consider the cases where $a_{31}a_{32}a_{21}\ne 1$. In these cases, an nonempty complete set consists of at most two elements. For each complete set $\mathcal{K}$, define the utility functions
\renewcommand\arraystretch{1}
\begin{equation}
u_{\mathcal{K}}(z)=
\left\{ \begin{array}{l}
\frac{\mu_i}{N_i+z}+\frac{[\mu_j-\mu_i]^+}{N_j+z},~\mathcal{K}=\{i,j\},i<j
\\
\frac{\mu_i}{N_i+z},~\mathcal{K}=\{i\},
\end{array} \right.
\end{equation}
and $u_+(z)=\max_{\mathcal{K}}u_{\mathcal{K}}(z)$, where $x^+\equiv \max(x,0)$. Denote $J^*$ as the optimal value of problem \eqref{optimization}. Then we have $J^*=\int_0^Pu_+(z)dz$, where the solution of \eqref{optimization} can be determined by intersection points of $u_{\mathcal{K}}$ with $\mathcal{K}\in\mathcal{K}_{\boldsymbol{I}}$. Note that the intersection points are roots of a polynomial with degree $\le 2$. The remaining steps are to investigate these roots. We omit the details here due to the page limit.
\end{IEEEproof}
According to \eqref{maxweightedsumrate}, $\mathcal{R}_{in}$ is larger than the rate region achieved by time sharing, where network coding is adopted in complete sets.
\subsection{An Outer Bound on The Capacity Region}
In order to characterize the outer bound on the capacity region of the channel \eqref{gaussianbc}, we introduce the following definition.

\textbf{\emph{Definition 3:}} A receiver set $\mathcal{V}_1$ is a \emph{weaker} set of $\mathcal{V}_2$ if
 1) $\min\mathcal{V}_1 >\max\mathcal{V}_2$ and 2) $a_{ij}=0$ for $i\in\mathcal{V}_1$ and $j\in\mathcal{V}_2$.
 A \emph{degraded} set $\mathcal{D}_\mathcal{J}$ consists of receiver acyclic sets $D_j,j=1,$ $\ldots,J,$ such that each receiver set $D_j$ is a weaker set of $D_{j-1}$.

We establish an outer bound in the following theorem. The proof combines the techniques used in \cite{liu2009gaussian} and in \cite{bergmans1974simple}.

\textbf{\emph{Theorem 3:}} Any achievable rate tuple $(R_1,R_2,R_3)$ for the broadcast channel \eqref{gaussianbc} must satisfy
\begin{equation}\label{outerbound}
\sum_{k\in D_j}R_k\le C\Big( \frac{P_j}{\min_{i\in D_j}N_i+\sum_{m<j}P_m}\Big),~j=1,\ldots,J,
\end{equation}
for all degraded sets $\mathcal{D}_\mathcal{J}$,
where $\sum_{m=1}^JP_m=P$.
\begin{IEEEproof}
Fix set $\mathcal{D}_\mathcal{J}$. For each acyclic set $D_j\in\mathcal{D}_\mathcal{J}$, suppose that it is ordered by
$D_j=\{i_j^1,\ldots,i_j^{|D_j|}\}$ such that $a_{i_j^ki_j^n}=0$ for $k>n$.
According to the Fano's inequality,
\begin{equation}
H(W_i|Y_i,\overline{\mathcal{W}}_i)\le n\delta_n,~~~~i=1,2,3,
\end{equation}
where $\delta_n$ tends to zero as $n\rightarrow \infty$.

Let $i_j=\min D_j$. Then
\begin{equation}\label{zeroentropy}
\begin{split}
&H((W_{i_j^l})_{l\le |D_j|}|Y_{i_j},\mathcal{W}\backslash\{W_{i_n^k}\}_{k\le |D_n|,n\le j})\\
&\le \sum_{l=1}^{|D_j|}H(W_{i_j^l}|Y_{i_j},\{W_{i_j^k}\}_{k>l},\mathcal{W}\backslash\{W_{i_n^k}\}_{k\le |D_n|,n\le j})\\
&\le \sum_{l=1}^{|D_j|}H(W_{i_j^l}|Y_{i_j^l},\overline{\mathcal{W}}_{i_j^l})\le|D_j|n\delta_n.
\end{split}
\end{equation}

Denote $\mathcal{W}'_j=\{W_{i_j^k}\}_{k\le |D_j|}$ and $\mathcal{W}'_0=\mathcal{W}\backslash\{\mathcal{W}'_j\}_{j\le J}$. Based on \eqref{zeroentropy}, we have
\begin{equation}\label{rateinequality}
\begin{split}
&n\sum_{k\in D_j}R_k=H(\mathcal{W}'_j)=H(\mathcal{W}'_j|Y_{i_j},\{\mathcal{W}'_k\}_{k>j},\mathcal{W}'_0)\\
&+I(\mathcal{W}'_j;Y_{i_j},\{\mathcal{W}'_k\}_{k>j},\mathcal{W}'_0)\\
&\le |D_j|n\delta_n 
+I(\mathcal{W}'_j;Y_{i_j}|\{\mathcal{W}'_k\}_{k>j},\mathcal{W}'_0)\\
&=|D_j|n\delta_n+H(Y_{i_j}|\{\mathcal{W}'_k\}_{k>j},\mathcal{W}'_0)\\
&-H(Y_{i_j}|\{\mathcal{W}'_k\}_{k\ge j},\mathcal{W}'_0).
\end{split}
\end{equation}
The rest of the proof follows by similar arguments of Bergman \cite{bergmans1974simple}. We briefly outline the remaining steps as follows.

For any achievable rate tuple $(R_1,R_2,R_3)$ such that
\begin{equation}
\sum_{k\in D_j}R_k= C\Big( \frac{P_j}{N_{i_j}+\sum_{m<j}P_m}\Big)+\delta'_j,~j=1,\ldots,J,
\end{equation}
where $\delta'_j\ge0$ for $j=1,\ldots,J$, and $\sum_{j=1}^J\delta'_j>0$. We have
\begin{equation}\label{22}
\begin{split}
&H(Y_{i_j}|\{\mathcal{W}'_k\}_{k>j},\mathcal{W}'_0)-H(Y_{i_j}|\{\mathcal{W}'_k\}_{k\ge j},\mathcal{W}'_0)\\
&\ge n(C\Big( \frac{P_j}{N_{i_j}+\sum_{m<j}P_m}\Big)+\delta'_j-|D_j|\delta_n),~j\le J.
\end{split}
\end{equation}
Define $g(S) \buildrel \Delta \over = \frac{1}{2}\ln(2\pi S)$. Notice that $\mathcal{D}_{\mathcal{J}}$ is degraded, we have $N_{i_{j+1}}>N_{i_j},~j=0,\ldots,J-1$. Based on the conditional entropy-power inequality in \cite{bergmans1974simple}, we obtain
\begin{equation}\label{23}
\begin{split}
&H(Y_{i_{j+1}}|\{\mathcal{W}'_k\}_{k\ge j+1},\mathcal{W}'_0)\\
&\ge ng(N_{i_{j+1}}-N_{i_j}+g(\frac{H(Y_{i_j}|\{\mathcal{W}'_k\}_{k>j},W'_0)}{n})^{-1}).
\end{split}
\end{equation}
Combining \eqref{22}-\eqref{23}, we can prove by mathematical induction that $H(Y_{i_J}|\mathcal{W}'_0)\ge ng(N_{i_J}+P)+n\sum_{j=1}^J\delta'_j-Jn\delta_n$. Since $\sum_{j=1}^J\delta'_j>0$, we obtain $H(Y_{i_J})>ng(N_{i_J}+P)$. This is a contradiction because $var(Y_{i_J})\le N_{i_J}+P$.
\end{IEEEproof}
\textbf{\emph{Remark 3:}} The conclusion in this theorem holds for multi-receiver channels under general side information configuration.



We now show that the inner and outer bounds are tight in some special cases.

\textbf{\emph{Theorem 4:}} For the Gaussian broadcast channel \eqref{gaussianbc}, the inner bound \eqref{capacityregion} and the outer bound \eqref{outerbound} are tight in the following cases: 1) $\mathcal{K}_{\boldsymbol{I}}=\{\{1,2,3\}\}$.
2) $\mathcal{K_{\boldsymbol{I}}}=\{\{k_1,k_2\},\{k_2,k_3\}\}$ ($k_1,k_2$, and $k_3$ are different) and $a_{k_1k_2}=a_{k_3k_2}=1$.
3) $\mathcal{K_{\boldsymbol{I}}}=\{\{k_1,k_2\},\{k_3\}\}$ ($k_1,k_2$, and $k_3$ are different) and $k_3 \ne 2$.
4) $\mathcal{K_{\boldsymbol{I}}}=\{\{1\},\{2\},\{3\}\}$.
\begin{IEEEproof}
By applying the results in Theorem 3 and Theorem 4, it can be proved that the capacity region is given by all rate tuples $(R_1,R_2,R_3)$ satisfying
\begin{equation}\label{tightcapacityregion}
\sum_{k\in \mathcal{V}}R_k\le\sum_{j=1}^M C\Big(\frac{P_j}{\min_{i\in\mathcal{K}_j\cap\mathcal{V}}N_i+\sum_{m< j}P_m}\Big)
\end{equation}
for all sets $\mathcal{V}$ such that $\mathcal{V}\cap\mathcal{K}_l$ is acyclic or empty. Here $M=|\mathcal{K}_{\boldsymbol{I}}|$, and $\mathcal{K}_l=\text{arg}\min_{\mathcal{K}\in\mathcal{\mathcal{K}_{\boldsymbol{I}}},\mathcal{K}\ni l}(\min\mathcal{K}+\max\mathcal{K})$.
\end{IEEEproof}
%

\textbf{\emph{Remark 4:}} The conditions listed in this theorem cover 46 out of all 64 possible message side information configurations.
%
%
%
%
%
%
%
%
%
%
%

To give more insight, consider the case when $\overline{\mathcal{W}}_1=W_3$, $\overline{\mathcal{W}}_2=\emptyset$ and $\overline{\mathcal{W}}_3=W_1$. The inner and outer bounds are not tight in this case. According to Theorem $2$, the inner bound \eqref{capacityregion} is the set of all tuples $(R_1,R_2,R_3)$ that satisfy
$R_1 \le C  \Big( \frac{P_3}{N_1+P_1+P_2} \Big) + C  \Big( \frac{P_1}{N_1} \Big),R_2 \le C  \Big( \frac{P_2}{N_2+P_1} \Big),R_3 \le C  \Big( \frac{P_3}{N_3+P_1+P_2} \Big) + C  \Big( \frac{P_1}{N_3} \Big)$
for some nonnegative $(P_1,P_2,P_3)$ such that $P_1+P_2+P_3=P$. Based on Theorem $3$, any rate tuple $(R_1,R_2,R_3)$ must satisfy
$R_1 \le C  \Big( \frac{P_1}{N_1} \Big),R_2 \le C  \Big( \frac{P_2}{N_2+P_1} \Big),$ and $
R_3 \le C  \Big( \frac{P-\frac{P_2N_2}{P_1+N_2}}{N_3+\frac{P_2N_2}{P_1+N_2}} \Big)$
for some tuple $(P_1,P_2)$ with $P_1+P_2\le P$. Fig.2 presents sample numerical results for the inner and outer bounds in this setting, where $P=10,N_1=0.2,N_2=0.5,$ and $N_3=1$. One can see that joint network and Gelfand-Pinsker coding achieves the capacity region within a small gap.
\begin{figure}\label{fig:2}
\centering
\includegraphics[width=0.46\textwidth]{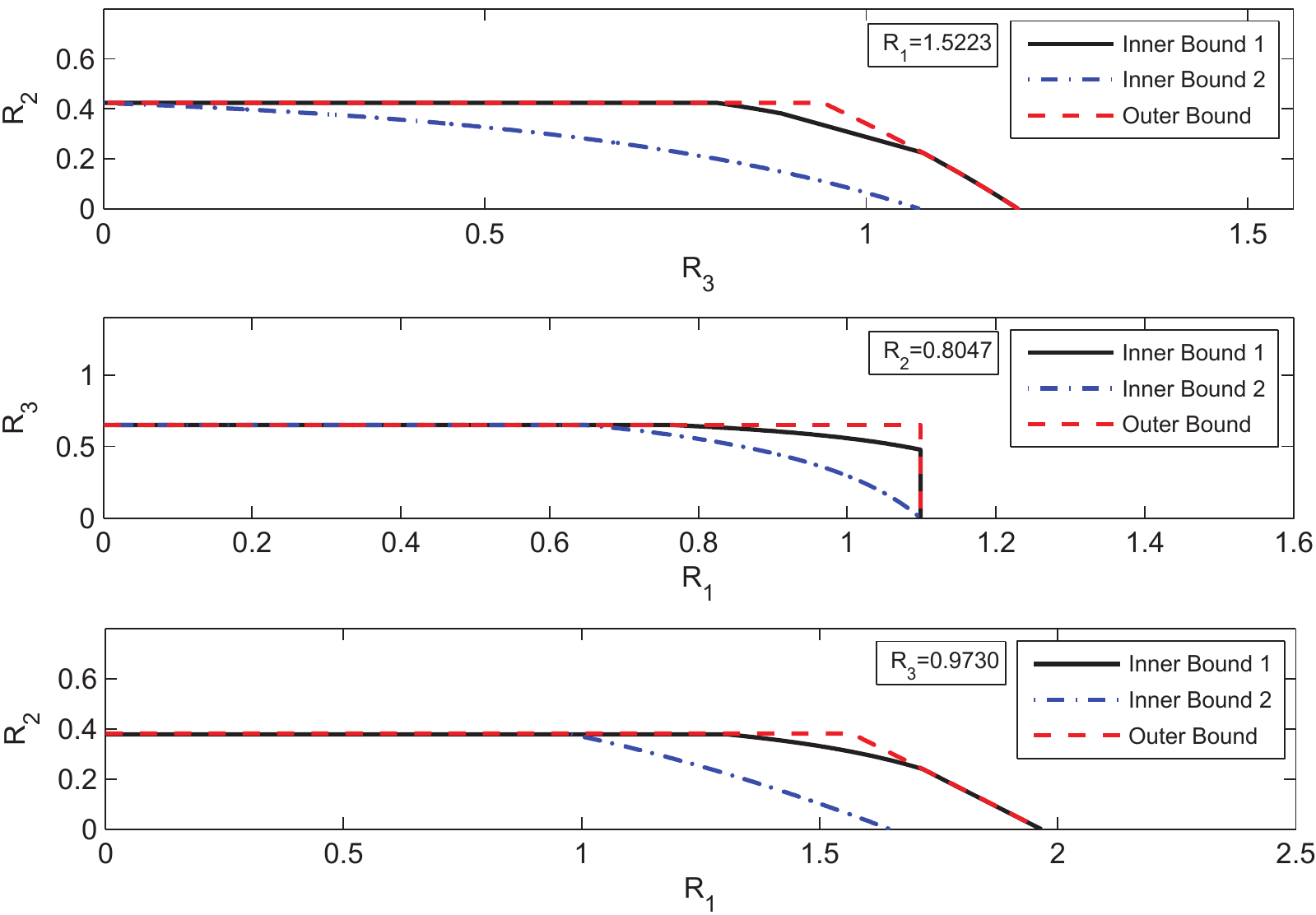} 
\caption{Inner and outer bounds in case $\mathcal{M}_1=W_3$, $\mathcal{M}_2=\emptyset$ and $\mathcal{M}_3=W_1$. Inner bound 1 is achieved by joint network and Gelfand-Pinsker coding and inner bound 2 is achieved by separate network and physical coding}
\setlength\belowcaptionskip{0pt}
\end{figure}

\section{Conclusion}
This paper we proposes joint network and Gelfand-Pinsker coding for 3-receiver Gaussian broadcast channels with receiver side information. The coding method provides a unified coding structure for general side information configurations. Using the proposed method and joint interference cancelation, we derive a unified inner bound to the capacity region of 3-receiver Gaussian broadcast channels under general side information configuration. The inner bound is shown to be larger than that achieved by state of the art coding schemes including time sharing and separate network and physical coding. We also present an outer bound on the capacity region and show that it is tight in 46 out of all 64 possible cases.


%
%

%
%

\bibliographystyle{IEEEtran}
\bibliography{IEEEabrv,confref}

\end{document}